\documentstyle[preprint,eqsecnum,aps]{revtex}
\begin{document}
\def\square{\kern1pt\vbox{\hrule height 1.2pt\hbox{\vrule width 1.2pt\hskip 3pt
   \vbox{\vskip 6pt}\hskip 3pt\vrule width 0.6pt}\hrule height 0.6pt}\kern1pt}

\def\lta{\mathrel{\spose{\lower 3pt\hbox{$\mathchar"218$}}
     \raise 2.0pt\hbox{$\mathchar"13C$}}}
\def\gta{\mathrel{\spose{\lower 3pt\hbox{$\mathchar"218$}}
     \raise 2.0pt\hbox{$\mathchar"13E$}}}
\def\spose#1{\hbox to 0pt{#1\hss}}
\def\om{\omega}
\def\cbr{cosmic background radiation}
\def\mpl{m_{\cal P}}

\newcommand{\be}{\begin{equation}}
\newcommand{\ba}{\begin{eqnarray}}
\newcommand{\ee}{\end{equation}}
\newcommand{\ea}{\end{eqnarray}}

\draft
\preprint{CITA-95-10}

\def\in{\hskip 5mm}

\in ${}$

\noindent{\bf GRAVITY DRIVEN INFLATION}

\hskip 0.8truein Janna J. Levin\par
{\baselineskip=14truept
\hskip 0.8truein Canadian Institute for Theoretical Astrophysics\par
\hskip 0.8truein McLennan Labs \par
\hskip 0.8truein 60 St. George Street\par
\hskip 0.8truein Toronto, ON  M5S 1A7\par

}
\vskip 2\baselineskip

\centerline{\bf Abstract}

The union of high-energy particle theories
and gravitation often gives rise to
an evolving
strength of gravity.
The standard picture of the earliest universe
would certainly deserve revision
if the Planck mass, which defines the
strength of gravity, varied.
A notable consequence is a gravity-driven,
kinetic inflation.
Unlike standard inflation, there is no potential
or cosmological constant.
The unique elasticity in the kinetic energy
of
the Planck mass provides a negative
pressure able to drive inflation.
As the kinetic energy
grows, the spacetime
expands more quickly.
The phenomenon of kinetic
inflation has been uncovered in both string theory
and
Kaluza-Klein theories.
The difficulty in exiting inflation in these cases
is reviewed.
General forms of the Planck field coupling
are shown to avoid the severity of the graceful exit
problem found in string and Kaluza-Klein theories.
The completion of the model is foreshadowed with
a suggestion for a heating mechanism to
generate the hot soup of the
big bang.

\vskip 2\baselineskip
To appear in the conference proceedings,
``Unified Symmetry:  In the Small and the Large II''
(Eds:  B.N. Kursunoglu, S. Mintz, and A. Perlmutter).

\vfill\eject

The theory of inflation \cite{alan} has become the
new paradigm for the early universe.
The inflation, or accelerated growth of the universe,
provides a dynamical explanation for the observed
smoothness on large-scales, the apparent local flatness,
and the lack of undesirable monopole relics.
Further, seeds to catalyze the formation of
galactic
structures on large-scales are predicted.
Though inflation occurs in the first
fraction of a second, the industry of
designing inflation models aspires
to explain the universe we live in today;
that is, to explain the specific features in the \cbr \ and
the formation of structure.\par

For all of its successes, there remain
some imposing questions.  In particular, there is no
model of inflation predicted from a
fundamental theory.
It might be
hoped that quantum gravity
would provide such a prediction.
However,
the most successful attempt at unifying
gravity with particle theory to date, namely
string theory, is inconsistent with the standard inflationary
paradigm \cite{steinhardt}.
While one might be willing to
abandon strings or inflation,
the point is we are no closer
to a consistent union of the
evolution of our universe and
a fundamental field theory.
Though  superstrings may not survive as the fundamental
theory, some of the
salient features must.
By investigating elements common to strings
and other particle theories in a
cosmological context, our larger
view of physics can be tested.\par

A common element of many high-energy theories is a dynamical
Planck mass.
In the low-energy string action
the dilaton acts as
a dynamical Planck field and thus
supplants the fundamental constant of the Einstein
theory.
Outside of superstrings,
dynamical Planck fields are often generated
in particle theories.
Even simple
quantum corrections to a field theory in a curved spacetime will
contribute to a variable Planck field.
In a higher dimensional or Kaluza-Klein approach
\cite{kk},
the variable
Planck mass has a geometric interpretation.
It is related to the radius of the compact internal
dimensions.\par

 A completely new source of inflation is predicted in such
extensions of Einstein gravity,
as we showed in references
\cite{{un},{me}}.
The phenomenon is manifest in string cosmology but is not unique
to string theory.
The nonminimal coupling of the Planck
field to gravity
allows for an
unusual elasticity
associated with the kinetic energy of the field.
The kinetic energy density can grow with time,
fueling a more rapid
expansion of the universe, that is, an inflation.
It is worth stressing
that there is no  potential and no
cosmological constant.
Independently,
Gasperini and Veneziano
considered the specific case of superstrings
\cite{gv}.\par

 In standard inflation,
a potential energy density
drives an era of accelerated expansion.
The characteristic feature is the potential.
Previously, string theories were shown to interrupt
potential dominated inflation
\cite{steinhardt}.
The kinetic energy in the dilaton field overwhelmed
the potential energy.  As a result, standard
inflation
could not proceed unhindered.\par

 Since the Planck field can actually
drive inflation, in lieu of the potential,
string theory may not only be
compatible with, but
actually predicts inflation.
I found an analogous type of behavior in
Kaluza-Klein models of additional
spacetime dimensions \cite{highd}.  In vacuum, the shear
from contracting dimensions is able to drive
an inflation of a three-volume.  When the extra
dimensions are integrated over,
the scenario is
equivalent to a four-dimensional model of a dynamical
Planck field.\par

 The task at hand is to uncover a successful end
to the scenario.
Currently, both string cosmology \cite{ram}
and Kaluza-Klein cosmologies \cite{highd} are unable
to exit the inflationary phase.
Instead the universe is
ushered toward a future singularity.
The graceful exit problem is more serious than
the usual obstacle which plagues potential-driven
inflation.  In string cosmology, the inflationary
branch of solutions is totally distinct from the
branch of solutions which describes our universe
today.  There is no overlap.  A branch change is
needed to move from inflation to a more
temperate evolution.
The nature of the graceful exit will be elaborated on here.\par

 Fortunately, the graceful exit problem does not
plague all models of a Planck-driven inflation.
As will be described, there are entire families which are
able to both inflate and match onto an expanding
universe today.  A means by which to heat the
universe, thereby completing the model, will also be described.\par

The gravitational action can be written in generality
as
\setcounter{section}{1}
	\be
	A_{G} = \int
	d^4x {\sqrt{-g}\over 16 \pi} \left [
	{-\Phi}{\cal  R} +
	{\omega\over \Phi}
	\left (\partial\Phi \right )^2 \right ]
 	\ . 	\label{act}
	\ee
The fundamental Planck scale
of the Einstein theory is replaced by the field $\Phi=\mpl^2$
and ${\cal R}$ is the Ricci scalar.
The kinetic coupling constant which determines the theory,
$\omega(\Phi)$, is left general.  To condense the notation it
is worthwhile to introduce the parameter
	\be
	f(\Phi)\equiv (1+2\omega(\Phi)/3)^{1/2}
	\ \ .
	\ee
In string theory, the dilaton is described by the action (\ref{act})
with $\omega=-1$.  In a Kaluza-Klein model with $n>1$ contracting
dimensions, the radius of the internal dimensions obeys
the action (\ref{act}) with
$\omega=-1+1/n$.  Notice that in both cases it happens that $f<1$.\par

 In a flat, Friedman-Robertson-Walker universe, the Einstein
equation which determines the expansion rate of the universe is simply
	\be
	H^2={8\pi \over 3\Phi}
	\left (\rho_\Phi +\rho\right )
	\ \ ,
	\label{ee}
	\ee
where the undecorated $\rho$ represents the energy density
in everything but the Planck field.
The kinetic energy density in $\Phi$ is given by
	\be
	{8\pi \over 3\Phi}\rho_\Phi=	{2\omega\over 3}
	\left ({\dot \Phi \over \Phi}\right )^2-H\dot \Phi
	\ \ .
	\ee
As a consequence of the direct coupling of the Planck field
to gravity, $\rho_\Phi$ involves $H$ directly.
The pressure associated with this
kinetic energy is roughly a measure of the change in energy
with unit volue, $p_\Phi \sim -{dE_\Phi/dV}$.
As a result of the direct coupling to the Ricci scalar,
the kinetic energy acquires a unique elasticity.  In fact,
for certain couplings $f$, the kinetic energy can actually
grow leading to a negative pressure.  In full glory, the
pressure associated with the kinetic energy can be written
	\be
	{8\pi \over 3\Phi}p_\Phi=
	{2\over 3} \left ({\dot \Phi \over 2\Phi}\right )^2
	\left (1+\omega\pm f -2 {d\ln f\over d\ln \Phi}\right )
	\ \ . \label{pe}
	\ee
The origin of the two branches is discussed below.
As in standard Einstein gravity, a negative pressure
can lead to inflation.\par

 Inflation refers to an accelerated growth of the scale factor.
The scale factor is accelerated if
the following condition \cite{me} is satisfied:
	\be
	f\pm 1-{\Phi \over f^2}{df\over d\Phi} <0
	\ \ .\label{co}
	\ee
If $\omega$ is a negative constant
so that $f<1$, as is the case for
strings and Kaluza-Klein,
then
condition (\ref{co}) will only be satisfied for the $-$ sign branch.
The condition becomes $f-1<0$, which is automatically satisfied.
If $\omega(\Phi)$ is variable, then it is
possible to satisfy condition (\ref{co})
for the $+$ sign branch.  The branch taken
turns out to be important.\par

 The physical relevance of these two branches can be seen by
solving the
quadratic Einstein equation (\ref{ee}) for $H$
	\be
	H=-\left ({\dot \Phi\over 2 \Phi}\right )
	\pm \sqrt{ {f^2\over 4}\left ({\dot \Phi\over \Phi}\right )^2
	+{8\pi \over 3\Phi} \rho }
	\ \ .\label{branch}
	\ee
The two branches in condition (\ref{co}) and expression
(\ref{pe}) reflect these two
solutions of Einstein's equations.
For comparison, the standard model Hubble expansion is given by
	\be
	H_{\rm stand}=\pm \sqrt{ {8\pi \over 3M_o^2} \rho }
	\ \ .\label{se}
	\ee
The standard Einstein equation (\ref{se}) also
allows two branches, one
expanding ($+$) and one contracting ($-$).  The expanding branch is
singled out
as the physically relevant one.  In the case of a dynamical
Planck mass on the other hand, both branches can expand
if $\dot \Phi <0$ and $f(\Phi)<1$.
In fact, for both Kaluza-Klein and strings,
the $+$ branch expands without inflation while the $-$
branch inflates.\par

 The pathology of the $-$ branch can now be seen.
Even if a mechanism exists to stabilize
the Planck mass,
the universe would ultimately contract.
Today the universe is described
by $+$ branch solutions of the form (\ref{se}).
A branch change is needed
to connect smooly onto
our expanding cosmology.  To induce such a branch change
requires
negative energies so that the total energy density drops to zero.
Obviously this is no mean feat.\par

The cure for string theory may lie in higher order
stringy corrections to the Einstein equations.
A different tact can be suggested.
If we allow for a variable $\omega(\Phi)$, then the
hardship of the branch change
can be circumvented.  The branch which
inflates can also be the branch which smoothly connects onto
an expanding universe today.
Entire families of couplings can satisfy the inflationary
condition (\ref{co}) on the healthier, $+$ branch.
Some examples of toy models include
	\ba
	f_1(\Phi)&=&{\ln(\Phi/M_o^2)}\nonumber \\
	f_2(\Phi)&=&
	\left ({1\over 4\ln(M_o^2/\Phi)}\right )^{1/2}\nonumber \\
	f_3(\Phi)&=&{\Phi\over M_o^2}\nonumber \\
	f_4(\Phi)&=&{\Phi \over M_o^2-\Phi }
	\ \ .
	\ea
The question remains if any such coupling can be generated from
quantum corrections to a low-energy effective action
or non-perturbatively from a high-energy theory.\par

 An accelerated expansion alone does not an inflationary
model make.  The universe must expand enough
to envelop our entire observable universe.  Furthermore,
the universe must then heat up to restore the standard
hot big bang picture.
Gravitational particle production due to the gravity-driven
inflation may be able to heat the universe
\cite{bond}.\par

 During inflation quantum fluctuations in any
underlying field theory can be
amplified.  For
wavelengths well within the event horizon, the amplified
fluctuations can propogate as particles.
The gravitational field transfers energy
into the
virtual quantum field thereby creating a hot bath.
On wavelengths which exceed an event horizon, the mode
cannot propogate as a particle but instead contributes to
the fluctuation
in an inhomogeneous classical
background.  The long wavelength fluctuations are the
usual density perturbations generated during inflation.  The short
wavelength fluctutations are the analog of Hawking-Unruh radiation.\par

 In de Sitter inflation, the short wavelength fluctuations are redshifted
away as rapidly as they are generated.
The equilibrium of classical de Sitter is therefore
left undisturbed.  The main
contribution of these
high-frequency quantum fluctuations is to build up the long
wavelength $\delta \rho/\rho$ as the modes cross outside the
event horizon.
In kinetic inflation however, particle production  can
run away when the expansion is fiercest.
The back-reaction
of the spacetime in turn drains energy from the Planck field.  As the
Planck field slows, inflation would in principle be exited.\par

 A quick sketch of a gravity-driven, kinetic inflation unfolds
as follows.
The elastic nature of the kinetic energy in a variable
Planck mass can drive an epoch of inflationary expansion.
The kinetic energy in the Planck field can grow as inflation
proceeds.  Consequently, high-energy
physics becomes increasingly
important.
The final stage of a gravity-driven inflation will thus be
marked by the influence of quantum mechanics through
gravitational particle production.
It must still be shown that
conversion of the classical kinetic energy
into particles is
efficient enough to appease the
demands of successful inflation \cite{me2}.
If a hot universe can be created from this cold beginning,
a cohesive model of
gravity-driven inflation is within reach.\par

\centerline{\bf Acknowledgements}

Special gratitude to
J.R. Bond, R. Brustein,
N. Cornish, K. Freese and G. Starkman
for many valuable discussions.
I thank Behram Kursunoglu for
his continued efforts
in
organizing the Global
Foundation meetings
where this talk was given.
I am also grateful for the additional support of the
Jeffrey L. Bishop Fellowship.

\end{document}